# nsroot: Minimalist Process Isolation Tool Implemented With Linux Namespaces

Inge Alexander Raknes
Dept. of Chemistry
University of Tromsø – The Arctic University of Norway
Tromsø, Norway
inge.a.raknes@uit.no

Bjørn Fjukstad, Lars Ailo Bongo
Dept. of Computer Science
University of Tromsø – The Arctic University of Norway
Tromsø, Norway
bjorn.fjukstad@uit.no, larsab@cs.uit.no

*Abstract*—Data analyses in the life sciences are moving from tools run on a personal computer to services run on large computing platforms. This creates a need to package tools and dependencies for easy installation, configuration and deployment on distributed platforms. In addition, for secure execution there is a need for process isolation on a shared platform. Existing virtual machine and container technologies are often more complex than traditional Unix utilities, like chroot, and often require root privileges in order to set up or use. This is especially challenging on HPC systems where users typically do not have root access. We therefore present *nsroot*, a lightweight Linux namespaces based process isolation tool. It allows restricting the runtime environment of data analysis tools that may not have been designed with security as a top priority, in order to reduce the risk and consequences of security breaches, without requiring any special privileges. The codebase of *nsroot* is small, and it provides a command line interface similar to chroot. It can be used on all Linux kernels that implement user namespaces. In addition, we propose combining *nsroot* with the AppImage format for secure execution of packaged applications. *nsroot* is open sourced and available at: https://github.com/uit-no/nsroot

## I. Introduction

Container and virtual machine technologies are vital for secure and portable execution on computing platforms such as clouds. For scientific computing the use of container technologies, such as Docker [1], rkt (https://coreos.com/rkt/) and Snapcraft (http://snapcraft.io/), have the potential to make data analysis applications easier to install, port, and use. Specifically, containers allow packaging the application with its dependencies in a self-contained container that can easily be executed on a wide variety of computing platforms. However, container technologies rely on operating system services to provide the application with a name space and protection from other application. Hence, these must rely on operating system features that may not be available in all kernels and distributions, or they may require root access.

Our work is motivated by our experiences developing, deploying, and maintaining the META-pipe metagenomics data analysis service [2]. Such genomic data analysis is typically done through a collection of tools arranged in a pipeline where the output of one tool is the input to the next tool. The data transformations include file conversion, data cleaning, normalization, and data integration. A specific biological data analysis project often requires a deep workflow that combines 10-20 tools [3]. There are many libraries [4]–[6] with hundreds of tools, ranging from small, user-created scripts to large, complex applications [7]. Each tool has dependencies in form of libraries, but also data files such as specific versions of reference databases. These dependencies make it difficult to install, configure, and deploy a pipeline on a new computing platform. There are therefore recent initiatives, such as BioDocker (https://github.com/BioDocker) that aim to provide packaged genomics data analysis tools in containers.

The cost of producing genomics data is rapidly decreasing (http://www.genome.gov/sequencingcosts), so it has become necessary to move the analyses from the single server typically used to high performance clusters or clouds [2], [3], [8]. Since porting and maintaining a deep data analysis pipeline on a cluster or cloud is time-consuming and requires access to a large amount of compute and storage resources, it is increasingly common to provide a pipeline as a data analysis service. Recent examples of such services include Illumina Basespace (https://basespace.illumina.com/home/index), EBI Metagenomics [9], our META-pipe [2] and other pipelines provided by the Norwegian eInfrastructure for Life Sciences (https://nels.bioinfo.no/). For such services security is important since the service may be used to process sensitive, scientifically valuable, or commercially valuable data on behalf of a user. This is further complicated since the pipeline consists of third party tools that may not have high code quality, that have not been properly tested, and that are executed as the same user on the cloud or HPC backend. So it is especially important to isolate the individual pipeline tool executions from each other. Another challenge is that, unlike IaaS services, HPC systems are often provided as a traditional shared UNIX system where users typically do not have root access. This complicates the task of installing complex application dependencies.

An ideal container solution would be one that would let us package all of our application dependencies inside the same Jar-file that we submit to our Spark cluster while at the same time provide full process isolation for the $3^{rd}$ party tools that are run within the workflow. This would also make it easy to know which tools (and versions) were involved in processing a dataset since they would all be part of the application bundle. However, to our knowledge no existing solution existed that satisfies all needs. Existing virtual machine and container technologies are often more complex than traditional Unix utilities, like chroot, and often require root privileges.

In this paper we discuss our solution for minimalistic process isolation. It solves the problem of isolating an application from the rest of the filesystem and it also lets us restrict an application from communicating with other applications via network or IPC by using the network- and IPC namespaces. Our solution, called *nsroot* provides process isolation by making use of Linux namespaces in order to provide an isolated root file system for a process (similar to chroot) while at the same time cutting it off the network. This enables us to run applications with all their dependencies is a single directory while at the same time providing extra security through extra process isolation. In addition, we combine *nsroot* with AppImage container to provide a simple, packaged bundle for an application. Finally, we propose using *nsroot* with AppImage and package those inside a Java Jar file for easy submission to a Spark cluster.

*nsroot* works on Linux kernels newer than 3.10, and we have successfully tested it on the Fedora 22 and Ubuntu 14.04 LTS.

## II. NSROOT

*nsroot* is a process isolation tool that makes use of namespaces to allow an application to change the root directory. It is designed to be lightweight and minimalistic to ensure wide portability across Linux kernel versions and distributions. It can be used to run processes within their own virtual root file system, which in turn allows the caller full control over absolute paths that are referenced by the program.

Unlike other container tools, like Docker or LXC, nsroot does not require any installation. Instead it can be packaged as part of an application bundle.

### A. Usage

To use *nsroot*, we first create a new root file system containing the application and all its dependencies. The root filesystem can be created by for example setting up the application in a docker container and then using docker to export the application files and dependencies to the new root directory. Alternatively, the root file system can be manually created, or it may consist of a statically linked self-contained application. We also describe below how nsroot can be combined with AppImage to provide extra isolation for containers.

Second, we create a new directory for the old root filesystem in a subdirectory of the new root. This is needed by the PIVOT_ROOT system call (http://man7.org/linux/man-pages/man2/pivot_root.2.html) that changes the root directory.

Third, we call *nsroot* with the same parameters as chroot in order to change into the new filesystem. We can also specify bind-mounts in order to mount existing directories into this new root file system. To further restrict which processes the contained process can communicate with we can apply the network and IPC namespaces. We can validate that the network namespace is used by calling ifconfig inside the container.

### B. Namespaces

We implemented *nsroot* by using namespaces as described in the NAMESPACES man page (http://man7.org/linux/man-pages/man7/namespaces.7.html). In particular, we are using mount namespaces in order to change the mount table for a single process tree and we are using the user namespace in order to do so without requiring root access (https://www.toptal.com/linux/separation-anxiety-isolating-your-system-with-linux-namespaces). We also found a tutorial about mount namespaces useful for this project (https://www.ibm.com/developerworks/library/l-mount-namespaces).

As described in the man page [] "*User namespaces isolate security-related identifiers and attributes, in particular, user IDs and group IDs [...], the root directory, keys [...], and capabilities [...]. A process's user and group IDs can be different inside and outside a user namespace. In particular, a process can have a normal unprivileged user ID outside a user namespace while at the same time having a user ID of 0 inside the namespace; in other words, the process has full privileges for operations inside the user namespace, but is unprivileged for operations outside the namespace.*"

In other words, the process can run with root privileges in a user namespace without actually having root privileges elsewhere. Summarized, our code does the following:

1. Clone the process using the appropriate namespaces. This includes at least the user and mount namespaces (CLONE_NEWUSER, CLONE_NEWNS)

2. Subprocess bind-mounts the new root filesystem into a subdirectory in the new root (for example /path/to/newroot/mnt).

3. pivot_root is called in order to make newroot the new root filesystem. The old root filesystem becomes available under /mnt

4. User-specified shared directories are bind mounted from the old root filesystem (available under /mnt) into the new root file system.

5. The old root filesystem (/mnt) is unmounted

6. execvp is called in order to replace the current process image with an executable specified by the user.

### C. Implementation

*nsroot* is implemented in C and it consists of less than 500 lines of code. It has no dependencies other than the kernel. User namespaces were implemented in the Linux kernel version 3.8. We have tested *nsroot* with Linux kernels newer than 3.10 (Fedora 22 and Ubuntu 14.04). However, it does not currently work on CentOS since it does not yet support user namespaces (http://rhelblog.redhat.com/2015/07/07/whats-next-for-containers-user-namespaces/).

## III. FUTURE DIRECTIONS

We believe nsroot can be used to provide extra isolation for minimalistic container technologies. These can then be used to package and execute for example data analysis tools as Spark jobs. This combination makes it easier to port legacy genomics data analysis pipelines to the Spark framework. In this section we present our proposed solution.

### A. AppImage

AppImage (http://appimage.org/) is a technology for packaging applications for Linux. An AppImage application is a single ELF executable file that contains all of the application's dependencies and it is self-contained. *nsroot* can be used in an AppImage to provide extra isolation via Linux namespaces.

Using *nsroot* within an AppImage has several advantages. First, it provides software packaging to nsroot. Second, it allows to run less-than-secure AppImages in isolation. Third, *nsroot* removes the problem of absolute paths. Fourth, it increases security by stricter isolation, which is especially beneficial if the installed AppImage becomes old or unpatched.

The above observations are also shared by the *firejail* (https://github.com/netblue30/firejail) tool that is implemented similarly to *nsroot* and that can already be used with the AppImage format.

### B. Spark Pipelines

Bioinformatics workflows running on Apache Spark could benefit from using *nsroot* for isolating the execution of legacy 3$^{rd}$ party tools. In particular, we could package all of our application dependencies inside the same Jar-file that we submit to our Spark cluster using AppImage, and then use *nsroot* to provide full process isolation for the 3$^{rd}$ party tools that are run within the workflow.

### C. Teaching Tool

We believe that the minimalistic design, and the 420 lines code, of *nsroot* makes it useful as a teaching tool. For example, *nsroot* could be used in a project about containers in an operating systems course.

## IV. RELATED WORK

In this section we describe related work in process isolation tools based on Linux user namespaces and container technologies. In addition, we discuss how containers are used for genomics data analysis.

### A. nsjail and firejail

At the same time when we developed nsroot, other projects developed similar tools. nsjail (http://google.github.io/nsjail/) was started the day before nsroot, and firejail (https://firejail.wordpress.com/) had its first beta release in April 2014. Both *nsjail* and *firejail* uses Linux user namespaces, similarly to *nsroot*, for process isolation, and in addition they use seccomp-bpf to filter system calls. *nsjail* uses resource limits and cgroups to limit the CPU, memory usage, and address space sizes of applications. *nsjail* also provides virtual network inside the application (although this requires root access to setup). *firejail* natively supports the AppImage format.

The main advantage of *nsroot* over *nsjail* and *firejail* is that *nsroot* maintains compatibility with the chroot command and it has a smaller codebase (since it has fewer features). We believe this makes the command line interface easier to understand and use.

### B. Containers and Virtual Machines

Both Virtual machines (VMs) and containers are technologies used to provide an isolated environment for applications. They are both binaries that can be transferred between hosts. While a VM includes an application, its dependencies and an entire OS, containers share the underlying host OS and contain only the application and dependencies making them smaller in size. Containers are isolated processes in user space on the host OS. (as described in https://www.docker.com/what-docker).

In [10] Docker containers are shown to have better than, or equal performance as VMs. Both forms of virtualization techniques introduce overhead in I/O-intensive workloads, especially in VMs, but introduce negligible CPU and memory overhead.

For genomics pipelines the overhead of Docker containers will be negligible since these tend to be compute intensive and they typically run for several hours [11].

### C. Containers in Life Science Research

Recent projects propose to use containers for life science research. The BioDocker (https://github.com/BioDocker) and Bioboxes [12] projects addresses the challenge of installing bioinformatics data analysis tools by maintaining a repository of docker containers for commonly used data analysis tools.

Containers have also been proposed as a solution to improve experiment reproducibility, by ensuring that the data analysis tools are installed with the same responsibilities [13].

## V. CONCLUSION

We have presented our *nsroot* process isolation tool. It uses the Linux user namespaces to allow application processes to run with root privileges without actually having root privileges outside of the user namespace. We propose using nsroot for secure packaged application execution, and for Spark job distribution.

Our work was motivated by our experiences developing a genomics data analysis service that we have designed for cloud execution. We plan to use *nsroot*, and the proposed Spark solution to install, configure, deploy, and securely execute the data analysis tools for user submitted data using a single cloud user. Although we focused on genomics data analysis, we believe our solution can also be used for data analysis pipelines in other fields. *nsroot* and the proposed solutions are especially suited for deep pipelines with potentially insecure third party tools.

As future work we intend to implement the proposed integration with AppImage and the Java jar file deployment. We also plan to explore the possibility of providing versioned datasets to ensure complete reproducibility of a pipeline using for example Pachyderm (http://pachyderm.io/), which is a project that provides version control for data and a containerized pipeline system.

ACKNOWLEDGMENT

This work was done as part of the Norwegian E-infrastructure for Life Sciences (NeLS) project. This work was funded by the Research Council of Norway.

REFERENCES

[1] D. Merkel, "Docker: Lightweight Linux Containers for Consistent Development and Deployment," *Linux J*, vol. 2014, no. 239, Mar. 2014.
[2] E. M. Robertsen, T. Kahlke, I. A. Raknes, E. Pedersen, E. K. Semb, M. Ernstsen, L. A. Bongo, and N. P. Willassen, "META-pipe - Pipeline Annotation, Analysis and Visualization of Marine Metagenomic Sequence Data," *ArXiv160404103 Cs*, Apr. 2016.
[3] Y. Diao, R. Abhishek, and T. Bloom, "Building Highly-Optimized, Low-Latency Pipelines for Genomic Data Analysis," presented at the CIDR, 2015.
[4] J. Hillman-Jackson, D. Clements, D. Blankenberg, J. Taylor, A. Nekrutenko, and G. Team, "Using Galaxy to Perform Large-Scale Interactive Data Analyses," in *Current Protocols in Bioinformatics*, John Wiley & Sons, Inc., 2002.
[5] R. C. Gentleman, V. J. Carey, D. M. Bates, B. Bolstad, M. Dettling, S. Dudoit, B. Ellis, L. Gautier, Y. Ge, J. Gentry, K. Hornik, T. Hothorn, W. Huber, S. Iacus, R. Irizarry, F. Leisch, C. Li, M. Maechler, A. J. Rossini, G. Sawitzki, C. Smith, G. Smyth, L. Tierney, J. Y. H. Yang, and J. Zhang, "Bioconductor: open software development for computational biology and bioinformatics," *Genome Biol.*, vol. 5, no. 10, p. R80, 2004.
[6] J. E. Stajich, D. Block, K. Boulez, S. E. Brenner, S. A. Chervitz, C. Dagdigian, G. Fuellen, J. G. R. Gilbert, I. Korf, H. Lapp, H. Lehväslaiho, C. Matsalla, C. J. Mungall, B. I. Osborne, M. R. Pocock, P. Schattner, M. Senger, L. D. Stein, E. Stupka, M. D. Wilkinson, and E. Birney, "The Bioperl Toolkit: Perl Modules for the Life Sciences," *Genome Res.*, vol. 12, no. 10, pp. 1611–1618, Oct. 2002.
[7] J. Leipzig, "A review of bioinformatic pipeline frameworks," *Brief. Bioinform.*, p. bbw020, Mar. 2016.
[8] F. A. Nothaft, M. Massie, T. Danford, Z. Zhang, U. Laserson, C. Yeksigian, J. Kottalam, A. Ahuja, J. Hammerbacher, M. Linderman, M. J. Franklin, A. D. Joseph, and D. A. Patterson, "Rethinking Data-Intensive Science Using Scalable Analytics Systems," in *Proceedings of the 2015 ACM SIGMOD International Conference on Management of Data*, New York, NY, USA, 2015, pp. 631–646.
[9] A. Mitchell, F. Bucchini, G. Cochrane, H. Denise, P. ten Hoopen, M. Fraser, S. Pesseat, S. Potter, M. Scheremetjew, P. Sterk, and R. D. Finn, "EBI metagenomics in 2016 - an expanding and evolving resource for the analysis and archiving of metagenomic data," *Nucleic Acids Res.*, p. gkv1195, Nov. 2015.
[10] W. Felter, A. Ferreira, R. Rajamony, and J. Rubio, "An updated performance comparison of virtual machines and Linux containers," in *2015 IEEE International Symposium on Performance Analysis of Systems and Software (ISPASS)*, 2015, pp. 171–172.
[11] P. Di Tommaso, E. Palumbo, M. Chatzou, P. Prieto, M. L. Heuer, and C. Notredame, "The impact of Docker containers on the performance of genomic pipelines," *PeerJ*, vol. 3, p. e1273, Sep. 2015.
[12] P. Belmann, J. Dröge, A. Bremges, A. C. McHardy, A. Sczyrba, and M. D. Barton, "Bioboxes: standardised containers for interchangeable bioinformatics software," *GigaScience*, vol. 4, p. 47, 2015.
[13] C. Boettiger, "An Introduction to Docker for Reproducible Research," *SIGOPS Oper Syst Rev*, vol. 49, no. 1, pp. 71–79, Jan. 2015.